\documentclass[12pt]{article}
\usepackage{cite}
\usepackage{amsbsy}

\title{On the Nature of the Relativity Principle.}
\author{M. Toller \thanks{e-mail: toller@iol.it}\\ 
via Malfatti n. 8  \\
I-38100 Trento, Italy}
 
\begin{document} 
\maketitle
                 
\begin{abstract}

We discuss the nature and a general formulation of the relativity principle and we show that it can be justified starting from a strictly operational point of view. We give some remark on the connection with the spacetime symmetry groups. We also propose that the relativity principle can be considered as a definition of physics, that distinguishes it from the other scientific disciplines.

\bigskip


\end{abstract}

\section{Introduction.}

It has been emphasized by Poincar\'e \cite{Poinc,Poinc2} and it is now generally admitted, that the scientific statements contain a mixture of conventions, definitions and genuine informations about nature that permit previsions. In the present paper we discuss, from this point of view, the relativity principle.

We adopt a rather general formulation of this principle, namely the requirement that all the local frames (LFs) have to be considered {\it a priori} equivalent. It follows that the physical laws are independent of the velocity of the frame (the aspect that is usually emphasized), but also of its spatial orientation and of its location in space and time. Of course, we have to specify more clearly what we mean by ``LF'' and by ``{\it a priori}''.

For instance, in the framework of general relativity, the model we have in mind for a LF is an orthonormal tetrad of four-vectors (also called a {\it Vierbein}) in a tangent space at a point $x$ of the spacetime manifold $\mathcal{M}$. Note that the term ``local'' is interpreted in a very restrictive way, namely a LF contains only the amount of information which is strictly necessary to determine the value of an observable, for instance a component of the electric field. To specify only the point $x$ would not be sufficient. 

Since the tetrad is only defined at a given time, it is not meaningful to ask if a LF is inertial, accelerated or rotating. A tetrad defines uniquely a system of normal coordinates in a neighborhood of $x$, which represents a locally inertial frame. Another coordinate system, whith the same tangent vectors to the coordinate axes at the point $x$, could represent an accelerated or a rotating frame.

All the tetrads can be considered as the points of a differentiable manifold $\mathcal{S}$, a principal fibre bundle \cite{KN} with basis $\mathcal{M}$ and structural group $O^{\uparrow}(1, 3)$, the orthochronous Lorentz group. This group acts on $\mathcal{S}$ and defines an equivalence relation on this manifold. The equivalence classes, namely the orbits, are the fibres of the bundle and correspond to the points of $\mathcal{M}$, namely the events.

The physical meaning of the equivalence relation mentioned above is spacetime coincidence. Einstein has emphasized \cite{Einstein} that the absolute character of the spacetime coincidence is one of the fundamental assumptions of general relativity. Our discussion of the relativity principle is not restricted to the framework of general relativity and it does not need this assumption. It is also meaningful for a class of possible nonlocal theories in which a spacetime $\mathcal{M}$ is not defined. 

The idea that the space $\mathcal{S}$ of the LFs is more fundamental than the spacetime $\mathcal{M}$ and that spacetime translations and Lorentz transformations should be treated on an equal footing has been suggested in ref.\ \cite{Lurcat} and has been developed in several papers, see for instance ref.\ \cite{CSTVZ} for other references. 

\section{The relativity principle and the operational point of view.}

We start from an operational point of view \cite{Bridgman,Giles}, namely we assume that at least some of the objects (terms or relations) belonging to the mathematical structure underlying a physical theory have to be interpreted in terms of physical operations. Actually, we shall adopt a ``strict'' operational point of view, namely we stress that what is relevant is not a single concrete operation, but the set of rules, called a ``procedure'', contained in a specific document, which describes exhaustively how the operation has to be performed. For instance, in the description of a procedure it is not allowed to point the finger at some physical object.

In order to specify the spacetime conditions, namely where and when the operation is performed, the procedure must refer to some pre-existent physical object, which determines a LF \cite{Toller}. The simplest kinds of procedures are the ``measurement procedures'', which give a numerical result, and the ``transformation procedures'' which have the aim of building a LF, starting from a pre-existent one.

As we have seen above, a theory may consider a space $\mathcal{S}$, the points of which are interpreted as potential LFs. By ``potential'' we mean that not all of them need to be physically realized. It is important to remark, however, that the interpretation of these points cannot be strictly operational, since there is no procedure which can produce a new LF starting from nothing. A frame can only be ``pointed at'', but not defined operationally without a reference to another LF. 

It follows that a physical theory cannot privilege some LFs: all of them have to be treated, {\it a priori}, on an equal footing. Here {\it a priori} means before any measurement is performed. It is clear that the result of the measurement of some physical field can be used to discriminate between LFs. In conclusion, we may say that the relativity principle follows from the requirement that the theory has a strict operational interpretation.

A measurement procedure defines a scalar field on $\mathcal{S}$ and a transformation procedure defines a differentiable mapping of $\mathcal{S}$ into itself, called a {\it transformation}. Both the values of the scalar field and the transformation mapping may depend on the state of the system. If they do not depend on the state of the system, we say that they have a nondynamical character, or that they have no dynamical relevance. A one-parameter group of transformations is generated by a vector field on $\mathcal{S}$. Both the scalar fields, the transformations and the vector fields that generate transformations have a strict operational interpretation. 
 
A physical theory may also consider mathematical objects that have no direct operational interpretation \cite{Giles}, but sometimes a better theory can be obtained by eliminating these objects. This suggestion resembles Heisenberg's idea to eliminate unobservable quantities (the electron orbits) from atomic physics \cite{Heisenberg, Heisenberg2}. In fact, one can consider theories \cite{Toller} in which the manifolds $\mathcal{S}$ and $\mathcal{M}$ do not appear and are replaced by a semigroup of abstract transformations with a strict operational interpretation. In theories of this kind, the relativity principle is trivially satisfied. 

One can also present the relativity principle as the requirement that the geometric aspects of a theory can be reformulated only in terms of transformations, without any reference to the LFs and the events.

\section{Symmetry groups.}

The requirement that all the points of $\mathcal{S}$ are equivalent can be assured by a symmetry of the theory with respect to a group acting transitively on $\mathcal{S}$. A natural choice would be the group of all the diffeomorphisms of $\mathcal{S}$, but often the symmetry group is smaller . If it is too small, it does not act transitively on $\mathcal{S}$ and the validity of the relativity principle does not follow.

For instance, in a local theory the diffeomorphisms must respect the space-time coincidence, namely map fibres onto fibres. In general relativity there is a stronger limitation, namely the symmetry diffeomorphisms must commute with the transformations of the structural Lorentz group, namely they must be automorphisms of the structure of principal fibre bundle of $\mathcal{S}$. These automorphisms form a group that acts transitively on $\mathcal{S}$.

In special relativity the symmetry group is the orthochronous Poincar\'e group $\mathcal{P}$, which also acts transitively on $\mathcal{S}$. The symmetry group of classical mechanics, namely the Galilei group, has the same property.

One should not confuse the mappings defined by the transformations with the mappings belonging to the symmetry group. For instance, in special relativity some transformations generate the orthochronous Poincar\'e group $\mathcal{P}$, which acts freely and transitively on $\mathcal{S}$. This means that if we choose a fixed LF $\hat s \in \mathcal{S}$ every LF $s \in \mathcal{S}$ can uniquely be written in the form $s = g \hat s$ with $g \in \mathcal{P}$ and we can identify $\mathcal{S}$ and $\mathcal{P}$. The transformations are left translations of the group $\mathcal{P}$ and the symmetry mappings are right translations, which commute with the left translations \cite{Lurcat} and are interpreted as changes of the choice of the fixed element $\hat s$.

In a similar way, we can see that a fibre of the principal fibre bundle $\mathcal{S}$ can be identified with the manifold of the structural (Lorentz) group. The action of the structural group can be described as a left translation, while the right translations (depending on the fibre) are symmetry transformations.

All the symmetries described above do not affect the indices that label the components of the fields. In general, there are other symmetries which act linearly on these indices, as it is explained, for instance, in ref.\ \cite{Toller2}. 

It is important to remark that the symmetry group of a theory depends on its mathematical formulation and not only on its physical content. This is possible because we are also considering gauge symmetries \cite{HHKN,Hehl}. For instance, if we write special relativity using general coordinates, it has the same symmetry group as general relativity. One has to introduce a connection which has no dynamical relevance and to require that the Riemann curvature vanishes. General relativity is obtained by releasing this constraint and giving a dynamical meaning to the connection coefficients.

In a similar way \cite{CSTVZ}, one can formulate general relativity in terms of differential forms on $\mathcal{S}$ and obtain a formalism symmetric with respect to all the diffeomorphisms of  this manifold. In this case too, one introduces fields with no dynamical relevance. By giving them a dynamical meaning one could obtain new theories describing new long range fields \cite{CSTVZ}, but, in the absence of suggestions from experiments, this program has turned out to be rather difficult.

To enlarge a directly observed symmetry group, introducing at the same time new fields that explain why the enlarged symmetry is not observed, is a rather common practice in theoretical physics. We think that it is justified only if the new fields have other observable and observed consequences.

\section{A ``definition'' of physics.}

K. Popper \cite{Popper} introduced a ``demarcation criterion''  that permits one to distinguish science from other branchs of learning. It requires that scientific theories can, in principle, be disproved or falsified by experiments or empirical observations. 

It is clear that it is easier to falsify a theory if it has some kind of operational interpretation.  We think, however, that a strict operational interpretation is not a necessary prerequisite for falsificability. This means that some scientific theories may not have a strict operational interpretation and may not even satisfy the relativity principle.

We want to propose that the relativity principle, formulated in the preceding sections,  characterizes physics (including chemistry) with respect to the other scientific disciplines. This means that a scientific statement that does not satisfy the relativity principle does not belong to physics. For instance, when we say that the acceleration of a falling body at a given point of the earth is $g = 9.8 \, m s^{-2}$ we are not speaking of physics, but of geography. 

Physics deals with general laws that are supposed to hold with respect to an arbitrary LF, in particular at any place and time. Of course the physical laws have some limits of validity, but these limits can  involve, for instance, temperature or pressure, not the choice of the LF.

It is important to remember that, as it has been emphasized by Poincar\'e and Popper, the scientific theories have an hypothetic nature and they have to be corrected if they are empirically falsified. This is the way in which science advances. In particular, if a physical law does not agree with an experiment performed in an arbitrary LF, the theory is falsified and must be corrected. Instead, for instance, if one finds that some biological phenomena appear differently on a distant planet, there is no problem.

Although Aristotle wrote a book entitled ``Physics'', according to our rather restrictive definition physics did not exist before Galilei and Newton. In fact, before their work, one did not believe that terrestrial and celestial bodies followed the same mechanical laws. 

An excessive attention to the exact definition of the various branches of learning may be useless or even dangerous for the progress of knowledge. Our aim is only to clarify some concepts and not to rise barriers between different disciplines and even less to establish preferences between them.

For this reason, we cannot agree with the famous sentence attributed to E. Rutherford: ``All science is either physics or stamp collecting'' \cite{Ruth}. If one remarks that stamps are issued in a specific year and in a specific country, one may perhaps consider ``stamp collecting'' as a (very reductive) metaphor for the scientific disciplines that do not satisfy the relativity principle. With this interpretation (that unfortunately the author cannot confirm), Rutherford's sentence may agree with our proposal.

From a different point of view (that we do not share either), the universal character of the physical laws and a possible corresponding lack of ``concreteness'' may be considered as a drawback. For instance, the idealistic philosopher B. Croce \cite{Croce} claims that the natural sciences are a system of ``pseudo-concepts'', while the true concepts concern history, namely concrete events tied to a particular time and place. 

In conclusion, we suggest that the relativity principle is mainly a definition: the definition of physics. However, it implies very important informations about nature, namely that physics, defined in that way, exists. In other words, there are physical laws, claiming to hold in all the LFs, that presumably are not eternal, but survived a very large number of attempts to falsification for a rather long time. 

\section{Long range fields and the equivalence principle.}

Several authors have discussed recently the possible existence of privileged frames, in which the physical laws take a particular simple form (see for example refs.\ \cite{CK,CG}). In this case, the relativity principle and the Lorentz symmetry are violated, unless we introduce new dynamic fields, for instance a four-vector field such that the privileged frames are characterized by the vanishing of its three spatial components. One can rewrite the physical laws introducing the new field in such a way that they take the same form in all the LFs.  All the LFs are again {\it a priori} equivalent and the privilege is attributed to the value taken by the field.

A similar procedure can be adopted in other cases of violation of the relativity principle. As a trivial example, we remark that in a bounded terrestrial laboratory the vertical direction seems to be privileged, because bodies fall in this direction.  The relativity principle and the rotational symmetry are restored if we consider the gravitational field generated by the earth mass according to Newton's law.

Another kind of violation of the relativity principle arises if one finds that some physical ``constants'' actually depend on time. A varying gravitational coupling has been proposed a long time ago \cite{Dirac,Jordan,BD}, but it is not experimentally confirmed up to now. Observational indications for a varying fine structure constant, which is also an old idea \cite{Dirac,MTW}, have been given recently \cite{WMF}. The relativity principle is restored if the varying ``constants'' are considered as dynamical fields.

These examples and the considerations of section 3 seem to suggest that the relativity principle can always be restored by introducing new fields. As a consequence, it cannot be contradicted and it seems to be meaningless. However, it is more correct to say that the relativity principle is meaningful if complemented by a list of all the dynamical long range fields. Adding a field to the list is a nontrivial operation: one has to find the dynamical equations and the sources that determine the new field, their interaction with other fields, and so on. One obtains in this way a new more powerful theory and one may say that the relativity principle is a very useful guide for the advancement of physics.

It is expedient to add here a short remark concerning another founding principle of general relativity: the equivalence principle. In its simpler (weak) form it says that all the sufficiently small falling bodies move in the same way starting from the same initial conditions.

Of course, this is not true if the bodies are charged or polarizable and an electromagnetic field is present. One can specify that the body is electrically neutral, use suitable shields in order to decrease the electromagnetic field, or calculate the influence of the residual electromagnetic force and take the correspondig corrections into account. A (not yet observed) dependence on spin of the motion of a particle could be caused by the presence of a long range torsion field \cite{HHKN}. If one finds other deviations, one can try to save the equivalence principle by introducing other new long range fields.

The problem should be examined with more detail but it seems that the equivalence principle, also in its more elaborate forms \cite{MTW}, should be considered as a definition of the gravitational interaction, which permits one to distinguish it from the other long range forces. As the relativity principle, with which it has some overlap, it acquires a clear physical meaning only if it is complemented by a list of all the nongravitational long range fields.  

At present, the only well established long range fields are gravitation and electromagnetism, but, as we have observed above, there are several proposal for adding new items to the list. This is one of the most interesting problems of contemporary macroscopic experimental physics. From the theoretical point of view, it is interesting to find geometric interpretations of the new fields. A more difficult problem, not discussed here, is to find a microscopic quantum theory for them.

\end{document}